\documentclass[12pt]{article}
\usepackage[utf8]{inputenc}
\usepackage[english]{babel}
\usepackage{amsmath}
\numberwithin{equation}{section}
\usepackage{amsthm,amssymb}
\usepackage{bm}
\usepackage{amsfonts}
\usepackage[margin=1in,includefoot]{geometry}
\usepackage{fancyvrb}
\usepackage{dsfont}
\usepackage{mathtools}
\usepackage{mathrsfs}
\usepackage{float}
\usepackage{caption}
\usepackage{pgfgantt}
\usepackage{indentfirst}
\usepackage{tikz}
\usepackage[font=small,labelfont=bf]{caption}
\usepackage[nottoc]{tocbibind}
\usepackage[toc]{appendix}
\usepackage{algorithm}
\usepackage{multirow}
\usepackage{titlesec}
\usepackage[affil-it]{authblk}
\newcommand{\email}[1]{\texttt{\small #1}}
\allowdisplaybreaks

\newcommand{\E}{\mathbb{E}}

\newcommand{\1}{\mathbf{1}}

\newcommand{\V}{\mathbb{V}\text{ar}}
\newcommand{\C}{\text{Cov}}

\author[]{Yiping Guo\footnote{Email: \email{y246guo@uwaterloo.ca} (Yiping Guo)} }
\author[]{Johnny Siu-Hang Li\footnote{Email: \email{shli@uwaterloo.ca} (Johnny Siu-Hang Li)}}
\affil[]{\normalsize Department of Statistics and Actuarial Science, University of Waterloo}

\usepackage[round]{natbib}
\title{Kriging Methods for Modelling Spatial Basis Risk in Weather Index Insurances: A Technical Note}
\date{}

\allowdisplaybreaks

\begin{document}
\maketitle
\begin{abstract}
\noindent 
The use of weather index insurances is subject to spatial basis risk, which arises from the fact that the location of the user's risk exposure is not the same as the location of any of the weather stations where an index can be measured. To gauge the effectiveness of weather index insurances, spatial interpolation techniques such as kriging can be adopted to estimate the relevant weather index from observations taken at nearby locations. In this paper, we study the performance of various statistical methods, ranging from simple nearest neighbor to more advanced trans-Gaussian kriging, in spatial interpolations of daily precipitations with data obtained from the US National Oceanic and Atmospheric Administration. We also investigate how spatial interpolations should be implemented in practice when the insurance is linked to popular weather indexes including annual consecutive dry days ($CDD$) and maximum five-day precipitation in one month ($MFP$). It is found that although spatially interpolating the raw weather variables on a daily basis is more sophisticated and computationally demanding, it does not necessarily yield superior results compared to direct interpolations of $CDD$/$MFP$ on a yearly/monthly basis. This intriguing outcome can be explained by the statistical properties of the weather indexes and the underlying weather variables.


\end{abstract}


\section{Introduction}

Weather insurances are often used by farmers and agricultural firms to protect against themselves losses or damages incurred because of adverse, measurable weather conditions. Traditional weather insurance contracts are generally indemnity-based, meaning that their payoffs are based on the insured parties' actual losses. From the insurer's perspective, indemnity-based weather insurances entail relatively high administration costs and are subject to moral hazard \citep{Adverse-1993-Quiggin, Adverse-2014-Erhardt}. These problems must be factored into insurance prices, thereby affecting the affordability of weather insurances to the agricultural sector as a risk management tool. To mitigate the drawback of indemnity-based weather insurances, insurers may choose to offer weather index insurances, the payoffs of which are linked to certain weather indexes that are calculated on the basis of certain common weather variables such as temperature and precipitation. As weather indexes are objective, insurers do not need to validate reported losses from weather index insurances, thereby saving administration costs. The objectivity of weather indexes also reduces the risk of moral hazard.

As with other index-linked insurances, weather index insurances entail basis risk, the risk that the actual loss incurred by the insured is different than the payout from the policy \citep{Basis-2011-Dick}. In more detail, such basis risk is composed of two main components. The first component is structural basis risk, which arises from the imperfect relationship between the payoff function and the insured parties' actual losses. In practice, the payoff function is often an indicator function, so that an insured party receives a fixed amount of payoff if the weather index (e.g., maximum temperature or aggregate precipitation) to which the policy is linked exceeds a certain trigger level. However, the true underlying relationship between actual losses and the underlying weather index is much more complicated and unknown. Therefore, instead of an exact indemnification, weather index insurances are generally used for mitigating the uncertainty surrounding a target outcome (e.g., crop yields for farmers), and its effectiveness of such risk mitigation has been empirically evaluated in some literature based on data from different countries including Canada \citep{WD-2001-Turvey}, China \citep{WD-2014-Sun} and the United States \citep{WD-2018-Zhou}. 

The second component is \textit{spatial basis risk}, the risk that is investigated in this paper. Spatial basis risk exists because the coverage of weather stations is never perfect \citep{Basis-2013-Norton}. If the weather stations at which measurements of weather indexes are taken are too far from the location of the insured party's risk exposure, then the mismatch between payoff and actual loss is inevitably deepened. Spatial basis risk is of particular concern in the context of microinsurance, which is commonly seen in developing countries with a low density of weather stations \citep{Micro-2010-Hazell}. When the target location (location of the insured party's risk exposure) is considerably distanced from weather stations, it becomes necessary to estimate the relevant weather variables at the target location from the nearby observations. This important procedure is known as a \textit{spatial interpolation}. 

The most commonly adopted family of spatial interpolation methods is \textit{kriging}, which enables the user to statistically incorporate information from multiple locations into the prediction for the target location. Compared to the fields of geostatistics and spatial statistics, kriging techniques have been much less extensively studied in actuarial science and insurance, particularly in the context of weather risk management, even though they lend themselves very well to the modelling of spatial basis risk in weather index insurances. Notable previous studies of kriging techniques in the actuarial science and insurance domain include the work of \cite{Basis-2013-Norton} who adopt an empirical approach to study and quantify spatial basis risk that is inherent in weather index insurances using US data, the contribution of \cite{Interpolation-2019-Roznik} who compare different universal kriging and generalized additive models for interpolating daily temperature data in the context of agricultural microinsurance, and the paper by \cite{Spatial-2019-Boyd} who further study the impact of kriging daily temperature on spatial basis risk reduction by analyzing the correlation between estimated payoffs and reported forage yields. 

The literature reviewed in the previous paragraph has only studied temperature variables and their related indexes such as consecutive cooling days. However, apart from temperature, precipitation is also regarded as a crucially important weather variable by the agricultural sector; for example, \cite{Rainfall-1970-Murphy} shows that forage yields are heavily impacted by cumulative precipitations within certain time periods. Kriging techniques that perform satisfactorily for temperature data do not necessarily yield the same level of performance for precipitation data. This is because compared to distributions of temperatures, distributions of precipitations are typically heavily skewed and have a significant probability mass at zero. To fill this research gap, in this paper, we perform a deeper investigation of kriging techniques in the context of weather index insurance, with a focus on precipitations and their related indexes. 

We consider daily precipitations, as well as two precipitation indexes that are derived from daily precipitations. The two precipitation indexes we consider are (i) maximum precipitation per month in five consecutive days ($MFP$) and (ii) annual maximum consecutive dry days ($CDD$). Given how they are defined, $MFP$ and $CDD$ capture changes in the left and right tails of the underlying precipitation distributions. They are therefore well suited as bases of weather index insurances that aim to mitigate the risk associated with extreme weather events which may result in huge losses, for example, significant reductions in agricultural yield \citep{WD-2001-Turvey}. It is noteworthy that $MFP$ and $CDD$ are component indexes of the Actuaries Climate Index (ACI), co-developed by a number of professional actuarial organizations to help inform actuaries, public policymakers, and the general public about climate trends and some of the potential impacts of a changing climate.

The first objective of this paper is to compare the performance of a range of spatial interpolation methods in spatial interpolations of precipitations and precipitation indexes. We begin with simple methods including nearest neighbor and inverse distance weighting; then we consider kriging methods including standard ordinary kriging, universal kriging, and trans-Gaussian kriging. The performance of the candidate methods for different types of data, including raw daily precipitations and the two mentioned precipitation indexes, is gauged by cross-validated (CV) interpolation errors.  It is found that the optimal spatial interpolation methods for raw daily precipitations and the precipitation indexes are different, owing to the differences in their distributional properties.

Our second objective is to investigate how spatial interpolations of precipitation indexes $MFP$ and $CDD$ are best implemented in practice. To fix ideas, let us consider a farm owner who wishes to mitigate the uncertainty surrounding the yield of his farm by purchasing a weather index insurance that is linked to the $CDD$ applicable to the location at which his farm is located. However, the farm is distanced considerably from weather stations at which precipitation measurements can be taken. In this situation, there is a need to estimate the $CDD$ values at the farm's location (target location) with spatial interpolations. Generally speaking, there are two ways to implement such spatial interpolations. One way is to take a \textbf{direct approach} in which $CDD$ values at the target location are estimated by spatially interpolating $CDD$ values recorded at nearby weather stations directly. The direct approach can be implemented easily without tracking the raw precipitation observations, and is not computationally demanding as spatial interpolations are performed only on a yearly basis (the frequency at which $CDD$ is reported). Another way is to take a more sophisticated \textbf{two-stage approach}, in which we first, on a daily basis, spatially interpolate raw precipitations recorded at nearby weather stations to obtain an estimate of the raw precipitation at the target location every day, and then compute the $CDD$ values at the target location on the basis of the daily precipitation estimates obtained in the first stage. Intuitively speaking, the two-stage approach appears to incorporate more information into the resulting estimates. We perform numerical analysis to compare these two approaches. Interestingly, it is found that although the two-stage approach entails a heavier data requirement (as raw daily temperatures measured at all nearby stations are needed) and computationally demanding (as spatial interpolations have to be performed substantially more frequently), it does not produce any better prediction accuracy compared to direct interpolations of $CDD$/$MFP$ on a yearly/monthly basis. We provide a statistical argument to explain this intriguing finding, which may help insurers offering weather index insurances with their risk quantification and management processes.   

The remainder of the paper is organized as follows. Section 2 describes the data set used in the paper. Section 3 presents the five spatial interpolation methods we consider. Section 4 documents two numerical analyses. The first analysis evaluates the performance of the five spatial interpolation methods in the application to daily precipitation data. Through this analysis, differences between temperatures and precipitations in the context of spatial interpolation are highlighted. The second analysis compares the direct and two-stage approaches for spatially interpolating precipitation indexes $MFP$ and $CDD$. Finally, some concluding remarks are made in Section 5.


\section{Data Description and Visualization}

The data used in this paper originated from the US National Oceanic and Atmospheric Administration (NOAA) National Climatic Data Center, and is accessed through R package \textbf{STRBOOK}. Among various variables in the data set, we consider daily precipitation $P$ in millimeters (mm) and maximum temperature $T$ in $^{\circ}$F at 138 weather stations in the central USA, recorded between 1990 and 1993, as well as the latitude $Lat$ and longitude $Lon$ of each of the 138 weather stations. 

Figures 1 and 2 show the locations of the weather stations and the daily precipitations and maximum temperatures measured on some selected days. It can be seen that certain areas of the region have no weather station. For these areas, one may use spatial interpolation to estimate precipitations. In both figures, we observe a clustering phenomenon that daily precipitations and temperatures  at nearby observations tend to be similar. From Figure 1 we observe that daily precipitations vary significantly across the region, with some close-to-zero values and some extremely large values. This observation suggests that daily precipitations exhibit a strong non-normality, a statistical property that is well taken care of in the modelling work presented in the next sections. On the contrary, Figure 2 shows that daily temperatures are distributed much more regularly with few outliers. 

\begin{figure}[hbt!]
    \centering
    \includegraphics[width=15.5cm]{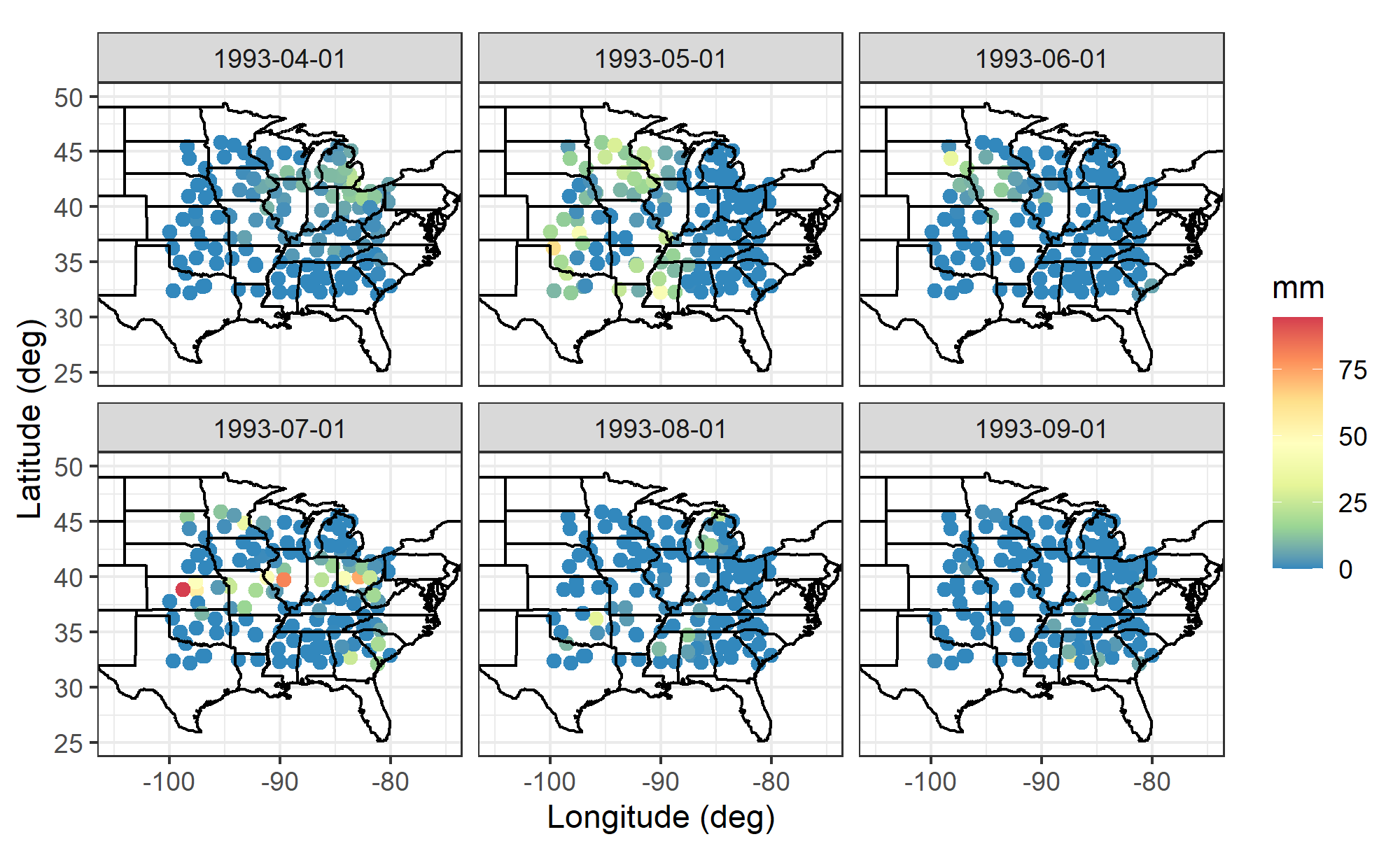}
    \caption{Daily precipitations $P$ (mm) from the NOAA data set on selected days in 1993}
    
    \vspace{0.25cm}
    \includegraphics[width=15.5cm]{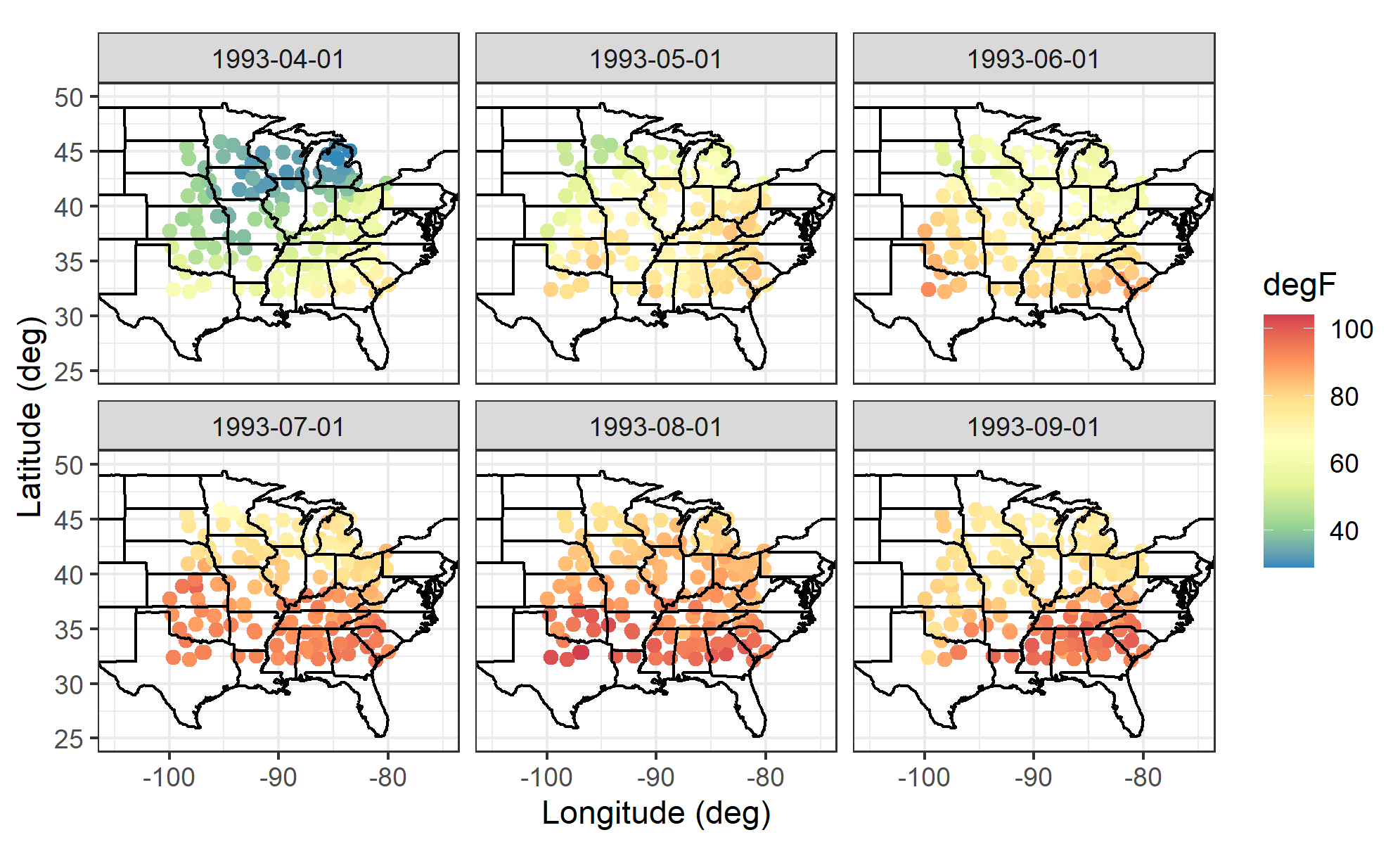}
    \caption{Daily maximum temperatures $T$ ($^{\circ}$F) from the NOAA data set on selected days in 1993}
\end{figure}

To further compare the two variables, in Figure 3 we plot seasonal histograms with estimated densities for daily precipitations and daily maximum temperatures. In line with the observations made in Figures 1 and 2, we observe from the histograms that daily precipitation data are extremely right-skewed and non-normal, whereas daily maximum temperatures are distributed fairly symmetrically in bell shapes. As elaborated in the next section, normality plays a crucial role in spatial interpolation, and therefore spatial interpolation techniques applicable to a certain weather variable may not be applicable to other weather variables without appropriate adaptations.

In addition to the four variables from the NOAA dataset, we consider the vertical elevation $Elev$ in meters of each weather station, as this variable is often taken into account as a covariate for interpolating precipitations in  climatology \citep{Elevation-1992-Phillips, Elevation-1996-Martínez-Cob} and actuarial science \citep{Spatial-2019-Boyd, Interpolation-2019-Roznik}. The elevation point data is obtained from the Elevation Point Query Service (EPQS) and the WGS84 coordinate system, through R package \textbf{elevatr}. 

\begin{figure}[t]
    \centering
    \includegraphics[width=16cm]{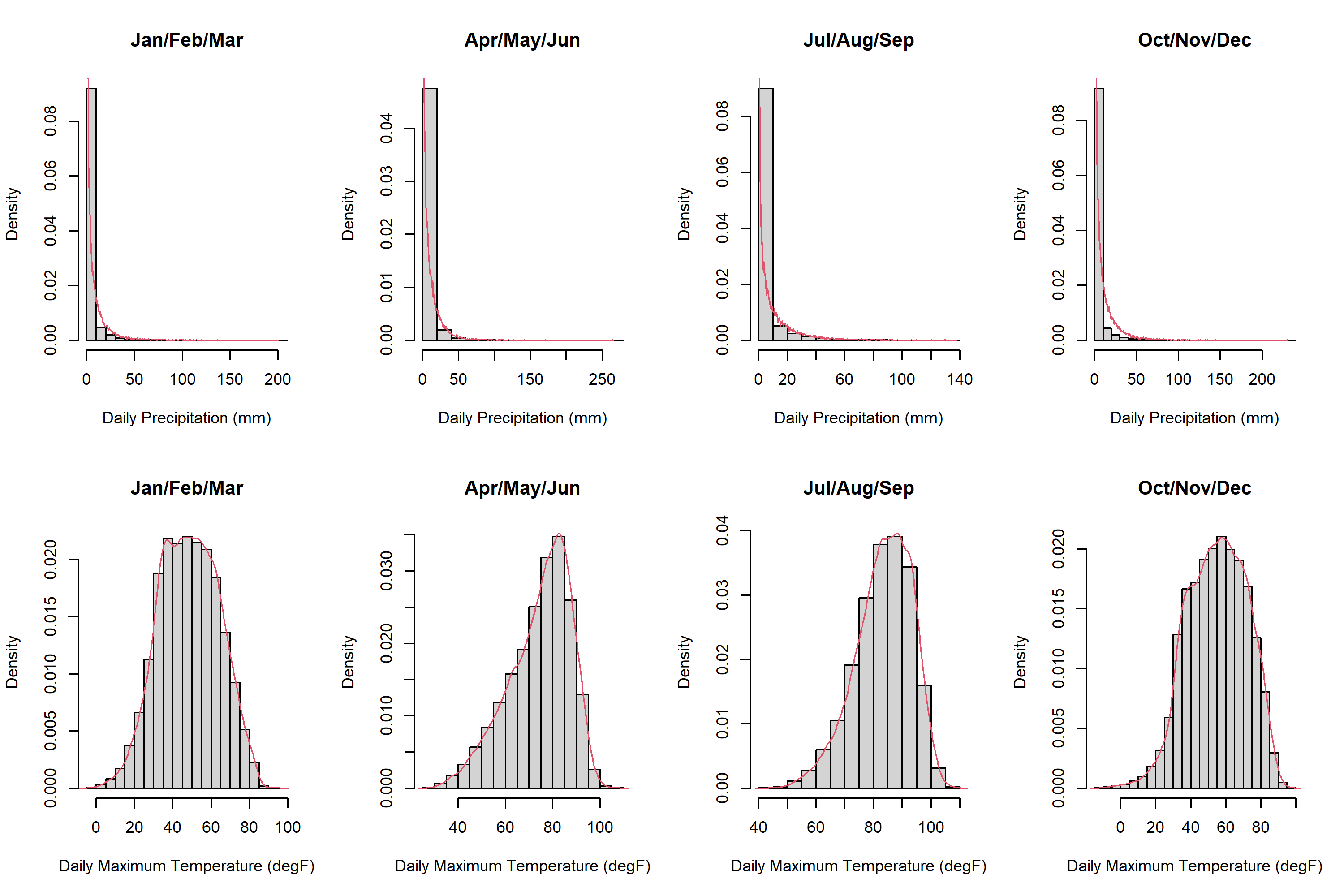}
    \caption{Histograms of daily precipitations $P$ (mm) and daily maximum temperatures $T$ ($^{\circ}$F) in different seasons}
\end{figure}

From the daily precipitation data, we calculate the historical values of two precipitation indexes on which weather insurances may be written. The first precipitation index we consider is the maximum precipitation per month in five consecutive days ($MFP$). The second precipitation index is the annual maximum consecutive dry days ($CDD$). To calculate $CDD$, we follow the definition adopted by  Actuaries Climate Index, which regards one day as a ``dry day" when the daily precipitation $P$ is below 1mm. 

For the reader's convenience, we summarize all of the variables defined earlier in Table 1: $Lat$, $Lon$ and $Elev$ are fixed covariates, $P$ and $T$ are daily observations, and $MFP$ and $CDD$ are precipitation indexes. 

\begin{table}[t]
\centering
\begin{tabular}{ccc}
\hline
Notation &  Name   & Data type   \\ \hline
$Lat$ & Latitude of the weather station & Continuous \\ 
$Lon$ & Longitude of the weather station & Continuous \\
$Elev$ & Elevation of the weather station & Continuous \\ \hline
$P$  & Daily precipitation (mm) & Continuous \\
$T$  & Daily maximum temperature ($^{\circ}$F) & Continuous  \\ \hline
$MFP$  & Maximum precipitation per month in five consecutive days & Continuous  \\ 
$CDD$  & Annual maximum consecutive dry days & Count \\
\hline
\end{tabular}
\caption{Variables defined in Section 2}
\end{table}



\section{Methodology}

In this section, we present the spatial interpolation techniques considered in this paper, including basic benchmark algorithms (nearest neighbor and inverse distance weighting), fundamental kriging methods (ordinary and universal kriging), and a more advanced kriging method known as trans-Gaussian kriging. We highlight the differences among these methods and discuss the appropriateness of these methods in the context of spatially interpolating precipitation-related quantities.  

Let the response variable and its estimate be $z(\bm{s})$ and $\hat{z}(\bm{s})$ in general, where $\bm{s}=(Lat,Lon)$ denotes the coordinate vector of the target location, with $Lat$ and $Lon$ representing the latitude and longitude, respectively. In our numerical analysis, $z$ can be either a basic weather variable like precipitation $P$ or a precipitation index like $CDD$. 

\subsection{Nearest Neighbor}
Nearest neighbor (NN) is the simplest spatial interpolation method, which directly takes the observation from the nearest weather station for the target location:
\begin{equation}\label{NN}
    \hat{z}(\bm{s}^{\ast})=z(\bm{s}_c),
\end{equation}
where $\bm{s}^{\ast}$ is the vector of the coordinates of the target location where a prediction is made, and $\bm{s}_c$ denotes the coordinate vector of the nearest weather station. Because of its simplicity and transparency, NN is easy to implement and understand and serves as a benchmark for evaluating the performance of more advanced spatial interpolation techniques. The drawback of this method is that it only takes the nearest point into account and ignores all information from other observations. It is therefore expected that NN produces relatively large interpolation errors.

\subsection{Inverse Distance Weighting}

Inverse distance weighting (IDW) generalizes NN by taking multiple nearby observations into account and calculating their weighted average as a prediction of the target location $\bm{s}^{\ast}$:
\begin{equation}\label{IDW}
    \hat{z}(\bm{s}^{\ast})=\sum_{i=1}^n w_i \cdot z(\bm{s}_i), 
\end{equation}
where $n$ is the number of nearby stations taken into consideration and 
\[w_i=\frac{\frac{1}{\Vert \bm{s}_i-\bm{s}^{\ast} \Vert^{p}} }{\sum_{i=1}^n \frac{1}{\Vert \bm{s}_i-\bm{s}^{\ast} \Vert^{p}}}\]
is the weight on location $i$ which reduces at a power rate as the distance $\Vert \bm{s}_i-\bm{s}^{\ast} \Vert$ between location $i$ and the target location increases. 

The pre-determined parameter $p$ controls rate at which the weight $w_i$ decreases with the distance $\Vert \bm{s}_i-\bm{s}^{\ast} \Vert$. A larger $p$ assigns more weight to closer points. We choose $p=2$ (square decay) as suggested by the classical geoscience literature \citep{IDW-2008-Li}. 

When applying IDW, it is a common practice to place certain arbitrary constraints on $n$ \citep{Spatial-2019-Boyd}. In this paper, we limit $n$ to 20, which means that we consider a maximum of 20 weather stations that are the closest to the target location.

The weights in (\ref{IDW}) sum to one, so that IDW estimators are unbiased if the underlying process of $z(\bm{s})$ has a constant mean over all locations. In fact, NN also enjoys the unbiasedness property since it is a special case of IDW when $n=1$. Therefore, NN and IDW are commonly adopted and used as benchmarks to evaluate the performance of more sophisticated spatial interpolation techniques. 

\subsection{Ordinary Kriging}

The main limitation of NN and IDW is that they restrict the form of spatial correlations to a power function. However, in practice, spatial correlations among observations are highly complicated and data-specific. Therefore, to produce more reliable spatial interpolations, it is necessary to model dependence structures via a more general family of functions which still retain the desirable properties of NN and IDW predictors such as linearity and unbiasedness. This necessity motivates the use of various kriging methods. This subsection focuses on ordinary kriging, which is foundational to the more sophisticated kriging methods discussed in the next two subsections. 

The formulation of ordinary kriging (OK) was originally proposed by \cite{Kriging-1951-Krige} and a more formal derivation of it was first provided by \cite{Kriging-1952-Davis}. The framework of ordinary kriging assumes that an observation $z(\bm{s})$ can be decomposed into an unknown stationary mean $\mu$ and a spatially correlated zero-mean noise $\varepsilon(\bm{s})$ as follows:
\begin{equation}\label{OK-model}
    z(\bm{s})=\mu+\varepsilon(\bm{s}).
\end{equation}
Then, for the target location $s^{\ast}$, OK aims to find the optimal unbiased predictor $\hat{z}(\bm{s}^{\ast})$, which takes the form of a homogeneously linear combination of other observations: $ \bm{z}=(z(\bm{s}_1),\cdots,z(\bm{s}_n) )$. Optimality is achieved by minimizing the mean squared prediction error subject to the unbiasedness condition: 
\begin{equation}\label{OK-MSE}
    \min_{\bm{\lambda}}\E[z(\bm{s}^{\ast})-\bm{\lambda}^T\bm{z}]^2 \quad \text{with }\bm{\lambda}^T\1=1,
\end{equation}
where $\bm{\lambda}=(\lambda_1,\cdots,\lambda_n)$ is the vector of kriging coefficients (kriging weights) to be determined, and $\1$ denotes a vector of ones. The solution can be obtained via the Lagrange multiplier method:
\begin{equation}\label{OK-solution}
    \hat{z}(\bm{s}^{\ast})=\underbrace{\hat{\mu}_{gls}}_{\text{mean estimator}} +\underbrace{\bm{c}(\bm{s}^{\ast})^T\bm{C}^{-1}}_{\text{weight}}\underbrace{(\bm{z}-\hat{\mu}_{gls}\cdot \1)}_{\text{detrended data}} \quad \text{with }\quad\hat{\mu}_{gls}=\frac{\1^T\bm{C}^{-1}\bm{z}}{\1^T\bm{C}^{-1}\1},
\end{equation}
where $\bm{C}=[\C(z(\bm{s}_i),z(\bm{s}_j))]_{i,j=1,\cdots,n}$ is the covariance matrix of the observations (made at nearby locations) and $\bm{c}(\bm{s}^{\ast})=(\C(z(\bm{s}_1),z(\bm{s}^{\ast})),\cdots,\C(z(\bm{s}_n),z(\bm{s}^{\ast})))$ is the vector of covariances between the predicted value and observations. Although the kriging predictor may be expressed in other equivalent forms such as the form of kriging equations \citep{Spatial-2015-Cressie},  (\ref{OK-solution}) is more interpretable. In (\ref{OK-solution}), the predictor $\hat{z}(\bm{s}^{\ast})$ is composed of two parts: (1) the trend term $\hat{\mu}_{gls}$ which represents the (restricted) generalized least squares estimate of the global mean $\mu$, and  (2) the mean ``correction'' term $\bm{c}(\bm{s}^{\ast})^T\bm{C}^{-1}(\bm{z}-\hat{\mu}_{gls}\cdot \1)$ that is expressed as a weighted sum of the detrended data $\bm{z}-\hat{\mu}_{gls}\cdot \1$, where the weights $\bm{c}(\bm{s}^{\ast})^T\bm{C}^{-1}$ depend on the spatial correlations.


In (\ref{OK-solution}), the covariances $\bm{C}$ and $\bm{c}(\bm{s}^{\ast})$ are taken as inputs, so they need to be estimated from the data $\bm{z}$. The principle behind spatial covariability modelling is that a specific family of covariance functions is fitted to the sample covariances, with a common assumption that $z(\bm{s})$ is second-order stationary, that is, $z(\bm{s})$ has a constant mean vector and the covariance between $z(\bm{s}_i)$ and $z(\bm{s}_j)$ for any $i \neq j$ depends only on the distance between  $\bm{s}_i$ and $\bm{s}_j$. Covariance functions $C(\bm{h})$ that are frequently used include Gaussian, exponential and spherical. They are all decreasing functions of the distance $\Vert \bm{h}\Vert=\Vert \bm{s}_i-\bm{s}_j \Vert$, but have different decaying rates. Following the classical literature in geostatistics for kriging rainfall variables \citep{Cov-2000-Goovaerts}, we choose the spherical covariance function:
\begin{equation}\label{Spherical}
    C(\bm{h})=\left\{ \begin{aligned}
&\sigma^2\left[1-\frac{3}{2}\cdot \frac{\Vert \bm{h}\Vert}{\alpha}+\frac{1}{2}\cdot \left(\frac{\Vert \bm{h}\Vert}{\alpha}\right)^3 \right], && 0\leq \Vert \bm{h}\Vert\leq \alpha\\
& 0, && \Vert \bm{h}\Vert>\alpha
\end{aligned}
\right.,
\end{equation}
where $\alpha$ is the practical range that allows the covariance vanishes if the distance between two observations becomes too large. The usual procedure for implementing OK is to first calculate the sample covariances of the observations, and then use the iterated generalized-least-squares (GLS) method \citep{Spatial-2015-Cressie} to estimate the parameters, which are $\sigma^2$ and $\alpha$ for the spherical case. For weather variables such as daily maximum temperature which are relatively regularly distributed, the covariance estimation procedure often works well and has a fast convergence rate. Nevertheless, for weather variables such as daily precipitations which feature highly imbalanced distributions with a large point mass at zero, the iteration might converge very slowly. This problem is practically important and is investigated deeper in the empirical analysis presented in Section 4.

The covariance fitting procedure is fairly well-developed and can be implemented with comprehensive geostatistical R packages such as \textbf{gstat} and \textbf{automap}. It is worth noting that most of these packages fit the so-called variograms instead of directly fitting the covariances. However, the end results of both routes are equivalent, because under the assumption of second-order stationarity, the variogram, defined as $2\gamma(\bm{h})=\V(z(\bm{s+h})-z(\bm{s}))$, has a one-to-one correspondence $2\gamma(\bm{h})=C(\bm{0})-C(\bm{h})$ with the covariance function $C(\bm{h})$ \citep{ST-2015-Cressie}. We choose to present OK in terms of covariance functions instead of the variograms, since covariance functions are more accessible to the finance community, and in fact, the optimization specified by (\ref{OK-MSE}) and (\ref{OK-solution}) is highly similar to a mean-variance portfolio optimization. 

\subsection{Universal Kriging}

The crucial underlying assumption behind OK is that the mean of $z(\bm{s})$ is constant over all locations and spatial dependence is completely captured by the residual term $\varepsilon(\bm{s})$. However, this assumption can be violated if the weather variable $z(\bm{s})$ is highly correlated with certain covariates; for example, daily maximum temperatures are usually strongly correlated with latitude. Such an effect should be removed before modelling the spatial correlation between the residuals \citep{UK-1994-Hudson}. One way to achieve this is to utilize universal kriging (UK), which assumes that $z(\bm{s})$ can be decomposed into a linear function of location-related covariates $\bm{x}(\bm{s})^T\bm{\beta}$ and a spatially correlated noise $\varepsilon(\bm{s})$:
\begin{equation}\label{UK-model}
    z(\bm{s})=\bm{x}(\bm{s})^T\bm{\beta}+\varepsilon(\bm{s}).
\end{equation}
The UK predictor takes a similar form to the OK predictor (\ref{OK-solution}). The only difference is that the UK predictor replaces the constant $\hat{\mu}_{gls}$ by a linear predictor $\bm{x}(\bm{s})^T\hat{\bm{\beta}}_{gls}$ \citep{ST-2015-Cressie}:
\begin{equation}\label{UK-solution}
    \hat{z}(\bm{s}^{\ast})=\underbrace{\bm{x}(\bm{s^{\ast}})^T\hat{\bm{\beta}}_{gls}}_{\text{mean estimator}} +\underbrace{\bm{c}(\bm{s}^{\ast})^T\bm{C}^{-1}}_{\text{weight}}\underbrace{(\bm{z}-\bm{X}\hat{\bm{\beta}}_{gls})}_{\text{detrended data}},
\end{equation}
where $\hat{\bm{\beta}}_{gls}=\left(\bm{X}^T\bm{C}^{-1}\bm{X}\right)^{-1}\bm{X}^T\bm{C}^{-1}\bm{z}$ is the GLS estimator of $\bm{\beta}$ and $\bm{X}=(\bm{x}(\bm{s}_1),\cdots,\bm{x}(\bm{s}_n))^T$ is the matrix of covariates at all locations. In this paper, we select the latitude, longitude, and elevation of a weather station as covariates for kriging precipitation-related quantities, that is, $\bm{x}=(Lat, Lon, Elev)$. 

As in ordinary kriging, the covariance matrix $\bm{C}$ and vector $\bm{c}(\bm{s}^{\ast})$ must be estimated. However, behind the same \textbf{R} function, the mechanism of fitting the covariance function is different. In OK the random part of the spatial dependence applies to the original data $\bm{z}$ since a stationary mean is assumed, whereas in UK the randomness comes from the residuals only. Thus, one must ``detrend'' the data $\bm{z}$ before modelling the spatial covariances. An ``optimal" detrending is not feasible, because the GLS estimate $\hat{\bm{\beta}}_{gls}$ of the trend involves the covariance matrix $\bm{C}$, which should be estimated after detrending. To circumvent this problem, one may obtain an initial estimate of $\bm{\beta}$ by a weighted least square (WLS) regression in which the weights might be chosen based on different rules \citep{Gstat-2004-Pebesma}; then, one can model the covariance function based on the detrended data and recalculate $\hat{\bm{\beta}}_{gls}$ for the final universal kriging predictor $\hat{z}(\bm{s}^{\ast})$.

It is documented in the literature that UK sometimes underperforms OK, even when the covariates for UK are empirically highly correlated to the interested quantity (e.g., latitude for daily maximum temperatures). This outcome is mainly caused by the intrinsically complex structure of the underlying spatial correlations, which are hardly driven by a handful number of covariates in a single linear form. Unless there exists a strong linear relationship between the chosen covariates and the observations, UK may yield an inferior covariance estimation and result in inaccurate kriging predictions.   

\subsection{Trans-Gaussian Kriging}

As shown from equation \eqref{OK-model} to \eqref{UK-solution}, the derivation of the OK and UK kriging predictors is based on the minimization of a mean squared prediction error, which does not depend on any specific distributional assumption. Despite this fact, it is important to note the relationship between kriging and Gaussian process regression. 

Gaussian process regression is a non-parametric approach which aims to determine the posterior distribution of the unobserved data given the observed data. In the derivation of the posterior mean, both observed and unobserved data are assumed to be drawn from a Gaussian process characterized by an unknown mean function and kernel. The posterior mean derived from Gaussian process regression and the kriging predictor derived from the minimization of a mean squared prediction error take the same form, which is represented by a weighted average of the observed values with the weights being determined by the underlying covariance structure. For this reason, the terms kriging and Gaussian process regression are sometimes used interchangeably, even though they are developed in different manners. The relationship between kriging and Gaussian process regression suggests that a kriging model is expected to yield superior prediction performance when the data follow closely to a Gaussian distribution. In other words, the normality of the underlying data does matter.

As shown in Figures 1 and 3, distributions of daily precipitations $P$ are non-Gaussian with heavy right tails, unlike the distributions of daily maximum temperatures $T$ which appear to be more Gaussian. To avoid large kriging predictive errors due to the non-normality of $P$, a simple strategy \citep{TG-2017-Cecinati} is to transform daily precipitation data with the Box-Cox transformation \citep{BC-1964-Box}: 
\begin{equation}\label{Box-Cox}
    y=\left\{ \begin{aligned}
&\frac{z^{\lambda}-1}{\lambda}, && \lambda \neq 0\\
& \log(z), && \lambda=0
\end{aligned}
\right.,
\end{equation}
where $y$ and $z$ represent the variables after and before transformation, respectively. The choice of the power parameter $\lambda$ relies on an empirical judgment. Following \cite{TG-2003-Sun}, we set $\lambda=1/3$ so that $y=3(\sqrt[3]{z}-1)$. It is worth noting that the Box-Cox transformation can only be applied to non-negative data, but this is not a concern in this study as precipitations are always non-negative. 

The approach of performing kriging algorithms on the transformed data $y$ instead of the original data $z$ is known as trans-Gaussian kriging (TGK). The implementation of TGK involves two stages. First, a standard non-transformed kriging algorithm is performed on the transformed data $y$. As such, a prediction $\hat{y}(\bm{s}^{\ast})$ for the target location $\bm{s}^{\ast}$ is obtained under the transformed scale. Second, a prediction $\hat{z}(\bm{s}^{\ast})$ of $z(\bm{s}^{\ast})$ in its original scale is made by back-transforming the transformed prediction $\hat{y}(\bm{s}^{\ast})$. 

In the second stage, a simple inverse transformation is inappropriate due to the fact that  $\E[z(\bm{s}^{\ast})]=\E[\phi (y(\bm{s}^{\ast}))] \neq \phi(\E[y(\bm{s}^{\ast})])$ for a non-linear function $\phi(\cdot )$. In this paper, we adopt the approximately unbiased estimator, obtained based on the delta method, recommended by \cite{Spatial-2015-Cressie}:
\begin{equation}\label{Back-transformation}
    \hat{z}(\bm{s}^{\ast})= \phi(\hat{y}(\bm{s}^{\ast}))+\phi^{\prime \prime}(\hat{\mu}_Y)\cdot \left(\frac{\sigma^2_{Y}(\bm{s}^{\ast})}{2}-m_Y \right),
\end{equation}
where $\phi(\cdot )$ is the inverse function of the chosen Box-Cox transformation, $\phi^{\prime \prime}(\cdot )$ is the corresponding second-order derivative, $\hat{\mu}_Y$ is the estimated mean defined in (\ref{OK-solution}), $\sigma^2_{Y}(\bm{s}^{\ast})$ is the kriging variance\footnote{The kriging variance is the minimized mean-squared prediction error, that is, $\E[z(\bm{s}^{\ast})-\hat{z}(\bm{s}^{\ast})]$, where $\hat{z}(\bm{s}^{\ast})$ is the kriging predictor. The exact formula can be found in classical spatial statistics texts \citep[e.g.][]{Spatial-2015-Cressie}.}, and $m_Y$ is the estimated Lagrange multiplier. We can obtain $\hat{\mu}_Y$, $\sigma^2_{Y}(\bm{s}^{\ast})$ and $m_Y$ from the standard ordinary kriging implementation. 

It is worth noting that trans-Gaussian kriging cannot be applied together with universal kriging. As such, in this paper, we only consider ordinary trans-Gaussian kriging for spatially interpolating daily precipitations. 


\section{Numerical Analysis}

In this section, we apply the spatial interpolation methods described in Section 3 to the NOAA data set, with the aim to answer the following two questions:
\begin{enumerate}
    \item Which of the spatial interpolation methods is the most appropriate for daily precipitations in terms of interpolation errors?
    \item When the weather index insurance under consideration is linked to $MFP/CDD$, would spatial interpolations of the raw precipitation data on a daily basis outperform those of $MFP/CDD$ itself on a monthly/yearly basis?
\end{enumerate}

Throughout the analysis, we measure predictive accuracy with a $K$-fold cross-validation (CV), an out-of-sample model validation technique that is widely used in geoscience \citep{CV-2008-Hofstra}, climatology \citep{CV-2010-Moral} and actuarial science \citep{Spatial-2019-Boyd, Interpolation-2019-Roznik} for comparing the performance of different spatial interpolation methods for weather variables. We implement CV with the following procedure. First, we randomly divide the observations to which a spatial interpolation is applied (e.g., daily precipitations recorded at the 138 weather stations on 1993-04-01) into $K$ equal-sized groups (folds). Second, For each of the $K$ groups, we predict the precipitation at every weather station in the group, on the basis of a spatial interpolation model that is fitted to the data from the remaining $K - 1$ groups. So, for each weather station with location $\bm{s}_i$, we have a predicted value $\hat{z}(\bm{s}_i)$ (obtained from the second step), which can be compared against its corresponding actual observed value $z(\bm{s}_i)$. Finally, the performance of the spatial interpolation is measured by the root mean squared error (RMSE), 
\begin{equation}\label{RMSE}
    \text{RMSE}=\sqrt{\frac{\sum_{i=1}^n(\hat{z}(\bm{s}_i)-z(\bm{s}_i))^2}{n}},
\end{equation}
and the mean absolute error (MAE),
\begin{equation}\label{MAE}
    \text{MAE}=\frac{\sum_{i=1}^n|\hat{z}(\bm{s}_i)-z(\bm{s}_i)|}{n},
\end{equation}
where $n$ denotes the sample size (the number of weather stations in our context). 

The choice of $K$, the number of folds, controls the balance between bias and variance. A smaller $K$ leads to a lower variance but a larger bias, whereas a larger $K$ results in the opposite. Following classic texts on model validation \citep{CV-1992-Breiman, CV-1995-Kohavi}, we choose $K=10$ to compromise.

The method of CV has been criticized by researchers such as  \cite{CV-2017-Roberts}, who argue that CV may underestimate predictive errors if the observations are not independent. The independence condition is clearly not satisfied in any spatial interpolation, which is by definition devised to capture the dependence of observations on the spatial domain. Thankfully, theoretical support for evaluating spatial interpolation methods with CV has recently been provided by \cite{CV-2022-Rabinowicz}, who rigorously formulate CV for dependent data and explicitly demonstrate the correctness of using CV in spatial interpolations. 

\subsection{Interpolating Daily Precipitations}


In this subsection, we utilize the previously discussed techniques to spatially interpolate daily temperatures from the NOAA dataset. Through the analysis, we can discern whether more advanced techniques such as UK and TGK can improve precipitation interpolation accuracy over simple benchmark techniques including NN and IDW. The results of this subsection are also useful in various means of weather risk analysis, for example, the creation of a high-resolution precipitation risk map that takes spatially interpolated daily precipitations as input. Further, the results in this section are relevant to our next analysis, which investigates whether spatially interpolating raw daily precipitation values may yield superior results compared to a direct interpolation of precipitation indexes such as $CDD$ and $MFP$ on a less frequent basis.

We apply NN, IDW, OK, UK and TGK to daily precipitations $P$ each day and calculate the corresponding RMSE and MAE with a 10-fold cross-validation. The resulting average RMSE and MAE over each year from 1990 to 1993 and the entire four year window are presented in Table 2.

\begin{table}[t]
\centering
\begin{tabular}{c c c c c c c}
\hline \hline
\multicolumn{7}{c}{RMSE ($P$)}\\
\hline
 Model &  Formula & 1990 & 1991 & 1992 &  1993  & 4-year average \\ \hline
NN  & N/A  & 
$5.93$ & $6.00$ & $5.44$ & $5.66$ & $5.76$ \\  
IDW & N/A  &
$4.86$ & $4.89$ & $4.47$ & $4.63$ & $4.71$ \\ 
OK  & $P\sim1$  & 
$4.87$ & $4.90$ & $4.51$ & $4.69$ & $4.74$\\ 
UK  & $P\sim Lat+Lon+Elev$ & 
$4.90$ & $4.93$ & $4.55$ & $4.73$ & $4.77$\\ 
TGK & $\sqrt[3]{P}\sim 1$ & 
$4.76$ & $4.82$ & $4.38$ & $4.60$ & $4.64$\\
\hline \hline
\\
\hline \hline
\multicolumn{7}{c}{MAE ($P$)}\\
\hline
 Model &  Formula & 1990 & 1991 & 1992 &  1993  & 4-year average \\ \hline
NN  & N/A             
& $2.49$ & $2.53$ & $2.33$ & $2.44$ & $2.44$ \\  
IDW & N/A                  
& $2.32$ & $2.31$ & $2.17$ & $2.26$ & $2.26$ \\ 
OK  & $P\sim1$             
& $2.45$ & $2.47$ & $2.32$ & $2.39$ & $2.41$\\ 
UK  & $P\sim Lat+Lon+Elev$ 
& $2.53$ & $2.54$ & $2.38$ & $2.47$ & $2.48$\\ 
TGK & $\sqrt[3]{P}\sim 1$  
& $2.07$ & $2.11$ & $1.97$ & $2.06$ & $2.05$\\
\hline \hline
\end{tabular}
\caption{Root mean squared errors (RMSE) and mean absolute errors (MAE) in the 10-fold cross-validations for different spatial interpolation methods applied to daily precipitations $P$. }
\end{table}

Let us first compare the two benchmark methods, NN and IDW. We observe that IDW produces significantly lower average RMSE and MAE compared to NN in each year and over the whole 4-year window. This result indicates that it is important to draw information from multiple nearby weather stations, and echoes the conclusions from previous studies \citep{Precip-2010-Chen, Precip-2015-Shope} that IDW generally serves as a better benchmark compared to NN. 

Before analyzing the results for more advanced spatial interpolation methods, let us make a practical note. When applying kriging methods to daily precipitations, it is important to consider the fact that distributions of daily precipitations are highly imbalanced with heavy right tails and a significant point mass at zero. The non-normality may cause potential issues when fitting a kriging model, particularly during the covariance function fitting stage. 
As previously mentioned, the parameters in the covariance function of the OK kriging predictor are estimated with a GLS iteration, which may converge slowly when non-normality is present. In the extreme scenario when all of the weather stations under consideration record zero precipitation on a day, then it is simply infeasible to fit any covariance model for the day and the GLS iteration will not converge. On the contrary, the solutions from NN and IDW always exist given their nonparametric formulations. To get around the possible non-convergence problem, we adopt IDW (which is demonstrated to perform better than NN) when an OK, UK or TGK fails to yield a converged estimate of the covariance function. 

Next, we turn to OK and UK. Although these techniques aim to capture spatial variability more precisely, in this application they underperform the benchmark method IDW in terms of both the RMSE and MAE. This result immediately raises the question as to whether it is necessary to consider more advanced kriging methods such as OK and UK. Further, this result seems to contradict some previous claims in the literature. For example, \cite{Spatial-2019-Boyd} and \cite{Interpolation-2019-Roznik} compare different spatial interpolation methods for mean daily temperatures and find that OK and UK generally produce lower RMSE compared to IDW. 

This seemingly anti-intuitive result can be attributed at least in part to the non-normality of daily precipitations. As demonstrated in Section 2, distributions of daily precipitations are far from Gaussian, whereas distributions of daily maximum temperatures are fairly close to normal. As normality is implicitly assumed in OK and UK, it is conceivable that they do not yield promising results when applied to daily precipitations which exhibit significant non-normality but perform satisfactorily when applied to daily maximum temperatures for which normality roughly holds.
Furthermore, we also observe that UK performs even worse than OK in this application. This result might be caused by the possibility that the underlying relationships between daily precipitations $P$ and the included covariates ($Lat$, $Lon$, and $Elev$) are non-linear so that the linear predictor in UK incorrectly detrend the observations and consequently introduce a bias when fitting the covariance function. 

Finally, we observe that TGK outperforms all of the other four methods (NN, IDW, OK, UK) consistently. Again, this result can be attributed to the non-normality of daily precipitations, a problem that is well handled by the transformation in TGK and the proper back-transformation specified in (\ref{Back-transformation}). Another interesting finding concerning TGK is that its improvement over the IDW benchmark is more significant in terms of (percentage reduction in) MAE than RMSE. This outcome is an indication that the improvement produced by TGK is mainly contributed by the better predictive quality for non-extreme points, as by definition (\ref{RMSE} and \ref{MAE}) RMSE penalizes large errors more heavily compared to MAE.  

To demonstrate the normalization effect of the Box-Cox transformation, we calculate the sample skewness and kurtosis of the original daily precipitations $P$ and their corresponding transformed values $\sqrt[3]{P}$. The same skewness and kurtosis calculations are also conducted for the daily maximum temperatures $T$, to illustrate the distributional differences between daily temperatures and precipitations. The definitions of sample skewness and sample kurtosis we adopt are as follows:
\begin{equation}\label{skewness}
\text{Skewness}={\frac {{\frac {1}{n}}\sum _{i=1}^{n}(x_{i}-{\overline {x}})^{3}}{\left({\tfrac {1}{n}}\sum _{i=1}^{n}(x_{i}-{\overline {x}})^{2}\right)^{3/2}}};
\end{equation}
\begin{equation}\label{kurtosis}
\text{Kurtosis}={\frac {{\tfrac {1}{n}}\sum _{i=1}^{n}(x_{i}-{\overline {x}})^{4}}{\left({\tfrac {1}{n}}\sum _{i=1}^{n}(x_{i}-{\overline {x}})^{2}\right)^{2}}}.
\end{equation}
In the above, $(x_1,\cdots,x_n)$ is the sample vector and $\overline{x}$ is the corresponding sample mean. The sample skewness is indicative of the symmetry of the underlying distribution, whereas the sample kurtosis reflects the heaviness of the tails of the underlying distribution. Samples from a normal distribution should have a sample skewness that is close to 0 and a sample kurtosis that is close to 3. We compute the sample skewness and kurtosis on a daily basis, and obtain the average values for each year from 1990 to 1993. The results are tabulated in Table 3. 

Before the transformation, daily precipitations $P$ has a sample skewness of 4.20 and a sample kurtosis of 26.83, which respectively suggest a significantly positive skewness and heavy tails. After transformation, the distribution of $\sqrt[3]{P}$ becomes closer to normal, thereby resulting in better kriging accuracy as depicted in Table 3. Arguably, the cube root transformation does not produce a perfect normality, as the resulting sample skewness (1.85) and kurtosis (8.26) are still quite different from the normality benchmark (0 for skewness and 3 for kurtosis). As a matter of empirical fact, the choice of transformation should not be based entirely on the proximity to normality after transformation. \cite{TG-2017-Cecinati} compare different Gaussian transformation methods for precipitation data, with a focus on selecting the optimal parameter of the Box-Cox transformation. Their results show that a transformation achieving the best normality does not necessarily result in the best kriging performance.  

On the other hand, the distribution of daily maximum temperatures is much closer to Gaussian, with an average sample skewness and kurtosis of $-0.08$ and $2.51$, respectively. This result offers an explanation as to why TGK is rarely considered when interpolating temperature-related variables. The proximity to normality also 
makes the non-convergence problem moot when fitting kriging models to temperature data.

\begin{table}[t]
\centering
\begin{tabular}{c c c c c c c}
\hline \hline
Metric & Variable & 1990 & 1991 & 1992 &  1993  & 4-year average \\ \hline
\multirow{3}{*}{Skewness}   
& $P$ & 
$4.32$ & $4.37$ & $4.01$ & $4.11$ & $4.20$ \\
& $\sqrt[3]{P}$ & 
$2.00$ & $1.96$ & $1.70$ & $1.74$ & $1.85$ \\ 
& $T$ & 
$-0.06$ & $-0.21$ & $-0.09$ & $0.03$ & $-0.08$ \\ 
\hline 
\multirow{3}{*}{Kurtosis}   
& $P$ & 
$28.22$ & $28.87$ & $24.58$ & $25.65$ & $26.83$ \\
& $\sqrt[3]{P}$ & 
$9.33$ & $9.19$ & $7.11$ & $7.43$ & $8.26$ \\ 
& $T$ & 
$2.47$ & $2.65$ & $2.47$ & $2.47$ & $2.51$ \\ 
\hline \hline
\end{tabular}
\caption{Average sample skewness and kurtosis for daily precipitations $P$, transformed daily precipitations $\sqrt[3]{P}$, and daily maximum temperatures $T$.}
\end{table}

Summing up, IDW serves as a reliable benchmark method as it consistently performs better than NN. In the application to daily precipitations, of which the underlying distribution deviates significantly from Gaussian, kriging methods outperform IDW only if a proper Gaussian transformation is adopted. There exist significant differences between kriging precipitations and temperatures (which are more normally distributed) in terms of both model selection and convergence issues, and therefore analysts should not transfer kriging approaches between different weather variables arbitrarily. 

\subsection{Interpolating Precipitation Indexes $MFP$ and $CDD$}

In this subsection, we focus on spatial interpolations for two precipitation indexes, maximum precipitation per month in five consecutive days ($MFP$) and annual maximum consecutive dry days ($CDD$), to which weather index insurances may be linked. From a practical perspective, there are two ways to spatially interpolate the two indexes:
\begin{enumerate}
    \item \textbf{Direct approach}: 
    
    In the direct approach, we perform spatial interpolations on $MFP$ and $CDD$ directly without considering the raw data from which the indexes are derived. This approach yields spatially interpolated values of $MFP$ every month and $CDD$ every year in one single step.
    
    \item \textbf{Two-stage approach}: 
    
    In the two-stage approach, we first perform spatial interpolations on the raw daily precipitations to obtain spatially interpolated precipitations at target locations every day. Then, following the definitions of $MFP$ and $CDD$, we compute the predicted values of $MFP$ every month and $CDD$ every year at target locations from the spatially interpolated daily precipitations obtained in the previous step.
    
\end{enumerate}
For each approach, we perform a 10-fold CV for the predicted values of $MFP$ every month and $CDD$ every year. These calculations result in, for each approach, an MAE every month and RMSE every year, which can be used to compare the performance of the direct and two-stage approaches in our application. The (average) values RMSE and MAE in each year from 1990 to 1993 and over the whole 4-year period, derived from both direct and two-stage approaches with the five spatial interpolation methods under consideration, are reported in Tables 4 (for $MFP$) and 5 (for $CDD$).

\begin{table}[t]
\centering
\begin{tabular}{c  c c c c c c}
\hline \hline
\multicolumn{7}{c}{RMSE ($MFP$)}\\
\hline
Approach & Model &  1990 & 1991 & 1992 &  1993  &  4-year average \\ \hline
\multirow{5}{*}{Two-Stage}   
& NN  &  
$28.23$ & $28.52$ & $23.93$ & $26.81$ & $26.87$ \\  
& IDW & 
$25.58$ & $25.78$ & $21.60$ & $23.12$ & $24.02$ \\ 
& OK  &  
$24.56$ & $25.29$ & $21.21$ & $22.37$ & $23.36$\\ 
& UK  &  
$24.11$ & $25.04$ & $20.93$ & $22.21$ & $23.07$\\ 
& TGK &  
$27.87$ & $28.22$ & $23.58$ & $25.01$ & $26.17$\\
\hline 
\multirow{5}{*}{Direct}
& NN               
& $28.23$ & $28.52$ & $23.93$ & $26.81$ & $26.87$ \\  
& IDW                   
& $23.91$ & $23.94$ & $19.88$ & $21.28$ & $22.25$ \\ 
& OK               
& $23.38$ & $23.70$ & $20.08$ & $21.60$ & $22.19$\\ 
& UK   
& $23.69$ & $24.12$ & $20.10$ & $21.81$ & $22.43$\\ 
& TGK   
& $22.87$ & $23.04$ & $19.67$ & $21.40$ & $21.75$\\
\hline \hline
\\

\hline  \hline
\multicolumn{7}{c}{MAE ($MFP$)}\\
\hline
Approach & Model &  1990 & 1991 & 1992 &  1993  &  4-year average \\ \hline
\multirow{5}{*}{Two-Stage}   
& NN  &  
$19.93$ & $19.54$ & $17.18$ & $19.01$ & $18.91$ \\  
& IDW & 
$17.18$ & $16.82$ & $14.91$ & $15.83$ & $16.19$ \\ 
& OK  &  
$16.64$ & $16.79$ & $15.01$ & $15.63$ & $16.01$\\ 
& UK  &  
$16.42$ & $16.54$ & $14.89$ & $15.62$ & $15.87$\\ 
& TGK &  
$18.58$ & $18.58$ & $16.33$ & $17.21$ & $17.68$\\
\hline 
\multirow{5}{*}{Direct}
& NN              
& $19.93$ & $19.54$ & $17.18$ & $19.01$ & $18.91$ \\  
& IDW                  
& $16.68$ & $16.32$ & $14.64$ & $15.27$ & $15.73$ \\ 
& OK              
& $16.37$ & $16.43$ & $14.84$ & $15.59$ & $15.81$\\ 
& UK  
& $16.87$ & $16.83$ & $14.86$ & $15.89$ & $16.11$\\ 
& TGK  
& $16.05$ & $15.74$ & $14.41$ & $15.33$ & $15.43$\\
\hline \hline

\end{tabular}
\caption{Root mean squared errors (RMSE) and mean absolute errors (MAE) calculated from the cross-validations of the spatial interpolations for $MFP$, implemented with the direct and two-stage approaches and different spatial interpolation methods.}
\end{table}

\begin{table}[t]
\centering
\begin{tabular}{c  c c c c c c}
\hline \hline
\multicolumn{7}{c}{RMSE ($CDD$)}\\
\hline
Approach & Model &  1990 & 1991 & 1992 &  1993  &  4-year average \\ \hline
\multirow{5}{*}{Two-Stage}   
& NN  & 
$6.63$ & $6.72$ & $6.03$ & $4.37$ & $5.94$ \\  
& IDW & 
$8.97$ & $10.53$ & $8.51$ & $6.73$ & $8.69$ \\ 
& OK  &  
$9.10$ & $12.70$ & $8.38$ & $7.54$ & $9.43$\\ 
& UK  &  
$9.25$ & $10.47$ & $7.87$ & $7.50$ & $8.77$\\ 
& TGK &  
$8.30$ & $11.32$ & $6.85$ & $6.36$ & $8.21$\\
\hline 
\multirow{5}{*}{Direct}
& NN               
& $6.63$ & $6.72$ & $6.03$ & $4.37$ & $5.94$ \\  
& IDW                   
& $5.41$ & $5.99$ & $4.63$ & $3.64$ & $4.92$ \\ 
& OK               
& $5.70$ & $5.87$ & $5.49$ & $3.47$ & $5.13$\\ 
& UK   
& $5.46$ & $6.19$ & $4.61$ & $3.43$ & $4.92$\\ 
& TGK   
& $5.58$ & $5.88$ & $5.61$ & $3.48$ & $5.14$\\
\hline \hline
\\

\hline  \hline
\multicolumn{7}{c}{MAE ($CDD$)}\\
\hline
Approach & Model  & 1990 & 1991 & 1992 &  1993  &  4-year average \\ \hline
\multirow{5}{*}{Two-Stage}   
& NN  & 
$4.16$ & $4.96$ & $3.98$ & $2.91$ & $4.00$ \\  
& IDW & 
$6.21$ & $7.84$ & $5.32$ & $4.30$ & $5.92$ \\ 
& OK  &  
$5.89$ & $9.22$ & $5.43$ & $4.84$ & $6.35$\\ 
& UK  &  
$6.44$ & $8.08$ & $5.33$ & $5.02$ & $6.22$\\ 
& TGK &  
$5.07$ & $7.72$ & $4.13$ & $3.71$ & $5.16$\\
\hline 
\multirow{5}{*}{Direct}
& NN  & $4.16$ & $4.96$ & $3.98$ & $2.91$ & $4.00$ \\  
& IDW & $3.51$ & $4.51$ & $3.34$ & $2.59$ & $3.49$ \\ 
& OK  & $3.79$ & $4.47$ & $3.76$ & $2.52$ & $3.64$\\ 
& UK  & $3.71$ & $4.89$ & $3.44$ & $2.48$ & $3.63$\\ 
& TGK & $3.72$ & $4.47$ & $3.83$ & $2.53$ & $3.64$\\
\hline \hline

\end{tabular}
\caption{Root mean squared errors (RMSE) and mean absolute errors (MAE) calculated from the cross-validations of the spatial interpolations for $CDD$, implemented with the direct and two-stage approaches and different spatial interpolation methods}
\end{table}

From Tables 4 and 5 we observe that for both $MFP$ and $CDD$, the direct approach produces smaller RMSE and MAE compared to the two-stage approach when the interpolation method used is IDW, OK, UK or TGK. The two approaches yield the same RMSE and MAE when the interpolation method used in NN, as NN makes use of the nearest observation to the target location only. The differences between the predictive errors resulting from the two approaches are particularly remarkable in the application to $CDD$.

In addition, Table 5 shows that the two-stage approach yields unreasonable results in the application to $CDD$: the predictive errors produced by IDW, OK, UK, and TGK are even higher than those from NN, which should not outperform the other four methods as we argued in Section 3. 

The striking results presented in Tables 4 and 5 beg explanations. The empirical fact that the two-stage approach under-performs the direct approach even though it is more sophisticated and computationally demanding can be attributed to the definitions (and hence statistical properties) of the  precipitation indexes.


For $MFP$, the two-stage approach first interpolates daily precipitations $P$. As the distribution of $P$ is not sufficiently close to Gaussian (even after the cube-root transformation in TGK), the kriging predictions of daily precipitations are generally not very satisfactory, and as a consequence, the calculated values of $MFP$ in the second stage might be inaccurate. In contrast, the direct approach directly interpolates $MFP$, the distributions of which are closer to normal. From Table 6 we observe that distributions of $MFP$ have an average skewness of 1.43 and average kurtosis of 6.06, suggesting that they are closer to Gaussian compared to the distributions of daily precipitations $P$ (with an average skewness of 4.20 and average kurtosis of 26.83; Table 3) and transformed daily precipitations $\sqrt[3]{P}$ (with an average skewness of 1.85 and average kurtosis of 8.26; Table 3). The higher proximity to normality can be attributed to an implicit smoothing effect introduced by the definition of $MFP$. As $MFP$ is calculated as the maximum precipitation per month in five consecutive days, as long as heavy precipitation days do not cluster, the distribution of $MFP$ should feature a less heavy right tail compared to that of $P$. 

For $CDD$, the index is computed as the longest run of dry days within a year through a ``rolling window'' approach, where a dry day is defined as a day with precipitation that is less than a strict threshold (the threshold used in this paper is $P\leq 1\text{mm}$). In the two-stage approach, the kriging errors in the first stage (where daily precipitations are spatially interpolated) can easily lead to a large number of misclassified dry days if some true precipitations are very close to the threshold. As just one single misclassification will break a run of dry days, the kriging errors in the first stage will ultimately result in highly inaccurate $CDD$ predictions.

Next, we compare the five spatial interpolation methods when the direct approach is taken. For $MFP$, TGK produces the most accurate predictions, a result that suggests that the Gaussian transformation remains important when spatially interpolating this participation index. However, compared to the application to $P$, TGK improves prediction errors over the benchmark method IDW only marginally, an outcome that might be attributed to the empirical fact that the distribution of $MFP$ is closer to normal compared to that of $P$ so that the benefit of the transformation is smaller.

\begin{table}[t]
\centering
\begin{tabular}{c c c c c c c}
\hline \hline
Metric & Variable & 1990 & 1991 & 1992 &  1993  &  4-year average \\ \hline
\multirow{4}{*}{Skewness}   
& $MFP$ & 
$1.53$ & $1.69$ & $1.24$ & $1.28$ & $1.43$ \\
& $\sqrt[3]{MFP}$ & 
$0.20$ & $0.25$ & $0.09$ & $0.09$ & $0.34$ \\ 
& $CDD$ & 
$1.59$ & $1.73$ & $1.83$ & $1.29$ & $1.61$ \\
& $\sqrt[3]{CDD}$ & 
$1.01$ & $1.08$ & $1.09$ & $0.77$ & $0.99$ \\ 
\hline 
\multirow{4}{*}{Kurtosis}   
& $MFP$ & 
$6.26$ & $7.44$ & $5.01$ & $5.54$ & $6.06$ \\
& $\sqrt[3]{MFP}$ & 
$3.27$ & $3.34$ & $3.27$ & $3.08$ & $3.24$ \\ 
& $CDD$ & 
$5.72$ & $6.00$ & $7.36$ & $4.93$ & $6.00$ \\
& $\sqrt[3]{CDD}$ & 
$3.64$ & $3.90$ & $4.69$ & $3.61$ & $3.96$ \\ 
\hline \hline
\end{tabular}
\caption{Sample skewness and kurtosis for $MFP$, $CDD$ and their transformed values.}
\end{table}

Interestingly, for the spatial interpolations of $CDD$ with the direct approach, TGK does not improve prediction accuracy over both OK and UK, even though $CDD$ features a similar degree of normality to $MFP$ as indicated by the sample skewness and kurtosis displayed in Table 6. Moreover, all of the three kriging methods (OK, UK and TKG) underperform the benchmark method IDW. These results are due possibly to a violation of the fundamental assumption of kriging  that an observation can be decomposed into a spatial trend plus a spatially correlated error term. To resolve this issue, one might consider more advanced nonlinear kriging methods such as multiple indicator kriging and probability kriging \citep{Spatial-2015-Cressie}, which are beyond the scope of this paper. 


\section{Concluding Remarks}

In this paper, we study a range of spatial interpolation methods for modelling spatial basis risk inherent in weather index insurances. Our empirical work is supported by weather data obtained from the NOAA in the United States. 

We extend the literature in the actuarial science and insurance domain by studying spatial interpolation methods for daily precipitations and precipitation indexes, which possess rather different distributional properties compared to temperature-related quantities that are considered in previous studies. For daily precipitations, we found that TGK is the best spatial interpolation method, an outcome that can be attributed to its Box-Cox transformation which largely eliminates the non-normality in daily temperatures. For precipitation index $MFP$, the conclusion remains the same but the improvement of TGK over the benchmark method is only marginal, because $MFP$ is somewhat more normally distributed compared to daily temperatures.

We also compare two approaches that may be taken to spatially interpolate precipitation indexes including $MFP$ and $CDD$ in practice: a direct approach in which $MFP$/$CDD$ are interpolated directly on a monthly/annual basis, and a two-stage approach in which the $MFP/CDD$ values at the target locations are computed from the estimated daily temperatures at the target location. To our knowledge, this study represents the first attempt to study this practically relevant problem. It is found that although the two-stage approach is more sophisticated and computationally demanding, it does not yield superior results compared to a direct interpolation of $CDD$/$MFP$ on a yearly/monthly basis. This intriguing outcome can be explained by the statistical properties of the precipitation indexes and their underlying weather variable.   

The findings of this study have direct implications on insurance. Our first key finding is that TGK reduces interpolation error compared to other candidate interpolation methods for $P$ and $MFP$. This finding suggests that the use of TGK may avoid an overestimation of the spatial basis risk in precipitation-related weather index insurances, and consequently allow a reduction in the risk-loading in the premiums of such insurances, thereby making such insurances more affordable. Our second key finding is that spatial interpolations for $CDD$ and $MFP$, both of which are components of the Actuaries Climate Index on which weather index insurances can be written, are better performed using a simple one-stage approach. This finding spares the insurance industry from unnecessary data collection and modelling work. We also believe that the desirable data-related property of the one-stage approach can be considered in tandem with the measures of effectiveness considered by \cite{Effective-2022-Pan} when evaluating weather indexes for risk management purposes.

Finally, it is found that for the spatial interpolations of $CDD$ with the direct approach, all of the three kriging methods (OK, UK and TKG) underperform the benchmark method IDW. As previously mentioned, this outcome is due possibly to a violation of the fundamental
assumption of kriging that an observation can be decomposed into a spatial trend plus a spatially correlated error term. In future research, it would be interesting to explore whether more advanced
nonlinear kriging methods such as multiple indicator kriging and probability kriging \citep{Spatial-2015-Cressie} can mitigate the unveiled issue concerning $CDD$.



\begin{thebibliography}{}

  \bibitem[\protect\citeauthoryear{Box and Cox}{Box and Cox}{1964}]{BC-1964-Box}
		Box, G. E., \& Cox, D. R. (1964). An analysis of transformations. \emph{Journal of the Royal Statistical Society: Series B (Methodological)}, 26(2), 211-243.
		
	\bibitem[\protect\citeauthoryear{Boyd et al.}{Boyd et al.}{2019}]{Spatial-2019-Boyd}	
		Boyd, M., Porth, B., Porth, L., \& Turenne, D. (2019). The impact of spatial interpolation techniques on spatial basis risk for weather insurance: An application to forage crops. \emph{North American Actuarial Journal}, 23(3), 412-433.	

  \bibitem[\protect\citeauthoryear{Breiman and Spector}{Breiman and Spector}{1992}]{CV-1992-Breiman}	
		Breiman, L., \& Spector, P. (1992). Submodel selection and evaluation in regression. The X-random case. \emph{International statistical review}, 291-319.

  \bibitem[\protect\citeauthoryear{Cecinati et al.}{Cecinati et al.}{2017}]{TG-2017-Cecinati}	
		Cecinati, F., Wani, O., \& Rico‐Ramirez, M. A. (2017). Comparing approaches to deal with non‐Gaussianity of rainfall data in kriging‐based radar‐gauge rainfall merging. \emph{Water Resources Research}, 53(11), 8999-9018.

  \bibitem[\protect\citeauthoryear{Chen et al.}{Chen et al.}{2010}]{Precip-2010-Chen}	
		Chen, D., Ou, T., Gong, L., Xu, C. Y., Li, W., Ho, C. H., \& Qian, W. (2010). Spatial interpolation of daily precipitation in China: 1951–2005. \emph{Advances in Atmospheric Sciences}, 27(6), 1221-1232.
  
    \bibitem[\protect\citeauthoryear{Cressie}{Cressie}{2015}]{Spatial-2015-Cressie}	
		Cressie, N. (2015). \emph{Statistics for spatial data}. John Wiley \& Sons.

   \bibitem[\protect\citeauthoryear{Cressie and Wikle}{Cressie and Wikle}{2015}]{ST-2015-Cressie}	
		Cressie, N., \& Wikle, C. K. (2015). \emph{Statistics for spatio-temporal data}. John Wiley \& Sons.

    \bibitem[\protect\citeauthoryear{Davis}{Davis}{1952}]{Kriging-1952-Davis}	
		Davis, R. C. (1952). On the theory of prediction of nonstationary stochastic processes. \emph{Journal of Applied Physics}, 23(9), 1047-1053.


  \bibitem[\protect\citeauthoryear{Dick et al.}{Dick et al.}{2011}]{Basis-2011-Dick}	
		Dick, W., Stoppa, A., Anderson, J., Coleman, E., \& Rispoli, F. (2011). Weather index-based insurance in agricultural development: A technical guide. \emph{International Fund for Agricultural Development (IFAD)}, 18.

  \bibitem[\protect\citeauthoryear{Erhardt and Smith}{Erhardt and Smith}{2014}]{Adverse-2014-Erhardt}
        Erhardt, R. J., \& Smith, R. L. (2014). Weather derivative risk measures for extreme events. \emph{North American Actuarial Journal}, 18(3), 379-393.
  



 \bibitem[\protect\citeauthoryear{Goovaerts}{Goovaerts}{2000}]{Cov-2000-Goovaerts}	
		Goovaerts, P. (2000). Geostatistical approaches for incorporating elevation into the spatial interpolation of rainfall. \emph{Journal of hydrology}, 228(1-2), 113-129.

  \bibitem[\protect\citeauthoryear{Hazell et al.}{Hazell et al.}{2010}]{Micro-2010-Hazell}	
		Hazell, P., Anderson, J., Balzer, N., Hastrup Clemmensen, A., Hess, U., \& Rispoli, F. (2010). The potential for scale and sustainability in weather index insurance for agriculture and rural livelihoods. \emph{World Food Programme (WFP)}.

  \bibitem[\protect\citeauthoryear{Hofstra et al.}{Hofstra et al.}{2008}]{CV-2008-Hofstra}	
		Hofstra, N., Haylock, M., New, M., Jones, P., \& Frei, C. (2008). Comparison of six methods for the interpolation of daily, European climate data. \emph{Journal of Geophysical Research: Atmospheres}, 228(1-2), 113(D21).


  \bibitem[\protect\citeauthoryear{Hudson and Wackernagel}{Hudson and Wackernagel}{1994}]{UK-1994-Hudson}	
		Hudson, G., \& Wackernagel, H. (1994). Mapping temperature using kriging with external drift: theory and an example from Scotland. \emph{International Journal of Climatology}, 14(1), 77-91.

  \bibitem[\protect\citeauthoryear{Kohavi}{Kohavi}{1995}]{CV-1995-Kohavi}	
		Kohavi, R. (1995). A study of cross-validation and bootstrap for accuracy estimation and model selection \emph{International Joint Conference on Artificial Intelligence (IJCAI)}, Morgan Kaufmann, 1137-1143.

      \bibitem[\protect\citeauthoryear{Krige}{Krige}{1951}]{Kriging-1951-Krige}	
	Krige, D. G. (1951). A statistical approach to some basic mine valuation problems on the Witwatersrand. \emph{Journal of the Southern African Institute of Mining and Metallurgy}, 52(6), 119-139.
  

    \bibitem[\protect\citeauthoryear{Li and Heap}{Li and Heap}{2008}]{IDW-2008-Li}	
	Li, J., \& Heap, A. D. (2008). A review of spatial interpolation methods for environmental scientists. \emph{Geoscience Australia}, Canberra, 2008.
		
	\bibitem[\protect\citeauthoryear{Martínez-Cob}{Martínez-Cob}{1996}]{Elevation-1996-Martínez-Cob}	
		Martínez-Cob, A. (1996). Multivariate geostatistical analysis of evapotranspiration and precipitation in mountainous terrain. \emph{Journal of Hydrology}, 174(1-2), 19-35.

  \bibitem[\protect\citeauthoryear{Moral}{Moral}{2010}]{CV-2010-Moral}	
		Moral, F. J. (2010). Comparison of different geostatistical approaches to map climate variables: application to precipitation. \emph{International Journal of Climatology: A Journal of the Royal Meteorological Society}, 30(4), 620-631.

  \bibitem[\protect\citeauthoryear{Murphy}{Murphy}{1970}]{Rainfall-1970-Murphy}	
		Murphy, A. H. (1970). Predicted forage yield based on fall precipitation in California annual grasslands. \emph{Rangeland Ecology $\&$ Management/Journal of Range Management Archives}, 23(5), 363-365.

  \bibitem[\protect\citeauthoryear{Murphy}{Murphy}{2012}]{ML-2012-Murphy}		
		Murphy, K. P. (2012). \emph{Machine learning: a probabilistic perspective}. MIT press.

  \bibitem[\protect\citeauthoryear{Norton et al.}{Norton et al.}{2013}]{Basis-2013-Norton}	
		Norton, M. T., Turvey, C., \& Osgood, D. (2013). Quantifying spatial basis risk for weather index insurance. \emph{The Journal of Risk Finance}, 14(1), 20-34.

 \bibitem[\protect\citeauthoryear{Pan et al.}{Pan et al.}{2022}]{Effective-2022-Pan}	
		Pan, Q., Porth, L., \& Li, H. (2022). Assessing the Effectiveness of the Actuaries Climate Index for Estimating the Impact of Extreme Weather on Crop Yield and Insurance Applications. \emph{Sustainability}, 14(11), 6916.

  \bibitem[\protect\citeauthoryear{Pebesma}{Pebesma}{2004}]{Gstat-2004-Pebesma}	
		Pebesma, E. J. (2004). Multivariable geostatistics in S: the gstat package.  \emph{Computers $\&$ geosciences}, 30(7), 683-691.
	
	
	\bibitem[\protect\citeauthoryear{Phillips et al.}{Phillips et al.}{1992}]{Elevation-1992-Phillips}	
		Phillips, D. L., Dolph, J., \& Marks, D. (1992). A comparison of geostatistical procedures for spatial analysis of precipitation in mountainous terrain. \emph{Agricultural and forest meteorology}, 58(1-2), 119-141.

  \bibitem[\protect\citeauthoryear{Quiggin et al.}{Quiggin et al.}{1993}]{Adverse-1993-Quiggin}	
		Quiggin, J. C., Karagiannis, G., \& Stanton, J. (1993). Crop insurance and crop production: an empirical study of moral hazard and adverse selection. \emph{Australian Journal of Agricultural Economics}, 37(429-2016-29192), 95-113.

  \bibitem[\protect\citeauthoryear{Rabinowicz and Rosset}{Rabinowicz and Rosset}{2022}]{CV-2022-Rabinowicz}	
		Rabinowicz, A., \& Rosset, S. (2022). Cross-validation for correlated data. \emph{Journal of the American Statistical Association}, 117(538), 718-731.

  \bibitem[\protect\citeauthoryear{Rasmussen and Williams}{Rabinowicz and Rosset}{2005}]{GP-2005-Rasmussen}	
		Rasmussen, C. E., \& Williams, C. K. I. (2005). \emph{Gaussian processes for machine learning}. MIT Press.

  \bibitem[\protect\citeauthoryear{Roberts et al.}{Roberts et al.}{2017}]{CV-2017-Roberts}	
		Roberts, D. R., Bahn, V., Ciuti, S., Boyce, M. S., Elith, J., Guillera‐Arroita, G., ... \& Dormann, C. F. (2017). Cross‐validation strategies for data with temporal, spatial, hierarchical, or phylogenetic structure. \emph{Ecography}, 40(8), 913-929.
		
	\bibitem[\protect\citeauthoryear{Roznik et al.}{Roznik et al.}{2019}]{Interpolation-2019-Roznik}	
		Roznik, M., Brock Porth, C., Porth, L., Boyd, M., \& Roznik, K. (2019). Improving agricultural microinsurance by applying universal kriging and generalised additive models for interpolation of mean daily temperature. \emph{The Geneva Papers on risk and Insurance-Issues and practice}, 44(3), 446-480.	

  \bibitem[\protect\citeauthoryear{Shope and Maharjan}{Shope and Maharjan}{2015}]{Precip-2015-Shope}	
		Shope, C. L., \& Maharjan, G. R. (2015). Modelling spatiotemporal precipitation: Effects of density, interpolation, and land use distribution. \emph{Advances in Meteorology}, 2015, 174196.

  \bibitem[\protect\citeauthoryear{Sun et al.}{Sun et al.}{2014}]{WD-2014-Sun}	
		Sun, B., Guo, C., \& van Kooten, G. C. (2014). Hedging weather risk for corn production in Northeastern China: The efficiency of weather-indexed insurance. \emph{Agricultural Finance Review}, 74(4), 555-572.

  \bibitem[\protect\citeauthoryear{Sun et al.}{Sun et al.}{2003}]{TG-2003-Sun}	
		Sun, X., Manton, M. J., \& Ebert, E. E. (2003). \emph{Regional rainfall estimation using double-kriging of raingauge and satellite observations}. Bureau of Meteorology.

  \bibitem[\protect\citeauthoryear{Turvey}{Turvey}{2001}]{WD-2001-Turvey}	
		Turvey, C. G. (2001). Weather derivatives for specific event risks in agriculture. \emph{Applied Economic Perspectives and Policy}, 23(2), 333-351.

  \bibitem[\protect\citeauthoryear{Zhou et al.}{Zhou et al.}{2018}]{WD-2018-Zhou}	
		Zhou, R., Li, J. S. H., \& Pai, J. (2018). Evaluating effectiveness of rainfall index insurance. \emph{Agricultural Finance Review}, 78(5), 611-625.
	 
\end{thebibliography}
\end{document}